\documentclass[twocolumn]{revtex4}

\usepackage{amsmath}
\usepackage{epsfig}
\usepackage{graphicx}
\usepackage{multirow}
\usepackage{algorithm}
\usepackage{algorithmicx}
\usepackage[noend]{algpseudocode} 

\newcommand{\argmax}{\operatornamewithlimits{argmax}}

\begin{document}

\title{Improving multivariate Horner schemes with Monte Carlo tree search}

\author{J.\ Kuipers}
\affiliation{Nikhef Theory Group, Science Park 105, 1098 XG Amsterdam, The Netherlands}

\author{A.\ Plaat}
\affiliation{Tilburg University, Tilburg center for Cognition and
  Communication, Warandelaan 2, 5037 AB Tilburg, The Netherlands}

\author{J.A.M.\ Vermaseren}
\affiliation{Nikhef Theory Group, Science Park 105, 1098 XG Amsterdam, The Netherlands}

\author{H.J.\ van den Herik}
\affiliation{Tilburg University, Tilburg center for Cognition and
  Communication, Warandelaan 2, 5037 AB Tilburg, The Netherlands}
 
\begin{abstract}
Optimizing the cost of evaluating a polynomial is a classic problem in
computer science. For polynomials in one variable, Horner's method
provides a scheme for producing a computationally efficient form. For
multivariate polynomials it is possible to generalize Horner's method,
but this leaves freedom in the order of the variables. Traditionally,
greedy schemes like most-occurring variable first are used. This
simple textbook algorithm has given remarkably efficient
results. Finding better algorithms has proved difficult. In trying to
improve upon the greedy scheme we have implemented Monte Carlo tree
search, a recent search method from the field of artificial
intelligence. This results in better Horner schemes and reduces the
cost of evaluating polynomials, sometimes by factors up to
two.
\end{abstract}

\maketitle

\section{Introduction}
Polynomials are fundamental objects in mathematics and reducing the
cost of evaluating polynomials is a classic problem in computer
science. Applications abound, ranging from fast calculation on
embedded devices and real-time calculations to high-energy physics
(HEP), where one needs to perform Monte Carlo integrations of
extremely large polynomials in many
variables~\cite{formcalc,grace,madgraph,gosam}. Numerous methods to
optimize polynomial evaluation have been proposed, such as Horner's
method~\cite{horner,ceberio,kojima}, common subexpression
elimination~\cite{compilers}, Breuer's growth
algorithm~\cite{breuer,hulzen} and, recently, partial syntactic
factorization~\cite{leiserson}.

For a polynomial in one variable, Horner's method provides a
computationally efficient form for evaluating it:
\begin{equation}
a(x) = \sum_{i=0}^n a_ix^i = a_0 + x (a_1 + x (a_2 + x(\dots + x\cdot a_n))).
\label{eqn::Horner}
\end{equation}
With this representation a dense polynomial of degree $n$ can be
evaluated with $n$ multiplications and $n$ additions, giving an
evaluation cost of $2n$. Here it is assumed that the cost of addition
and multiplication are equal.

For multivariate polynomials Horner's method can be generalized. To do
so one chooses a variable and applies Eqn.~(\ref{eqn::Horner}),
thereby treating the other variables as constants. Afterwards another
variable is chosen and the same process is applied to the terms within
the parentheses. This is repeated until all variables are
processed. As an example, for the polynomial
$a=y-3x+5xz+2x^2yz-3x^2y^2z+5x^2y^2z^2$ and the order $x<y<z$ this
results in the following expression
\begin{equation}
a = y+x(-3+5z+x(y(2z+y(z(-3+5z))))).
\label{eqn::Hornerexample}
\end{equation}
Regarding the evaluation cost, the original expression uses 5
additions and 18 multiplications, while the Horner form uses 5
additions but only 8 multiplications. In general, applying a Horner
scheme keeps the number of additions constant, but reduces the number
of multiplications.

After transforming a polynomial with Horner's method, the code can be
further improved by performing a common subexpression elimination
(CSE)~\cite{compilers}. In Eqn.~(\ref{eqn::Hornerexample}), the
subexpression $-3+5z$ appears twice. Eliminating the common
subexpression results in the code
\begin{equation}
\begin{array}{l}
T = -3+5z \\
a = y+x(T+x(y(2z+y(zT)))),
\end{array}
\end{equation}
which uses only 4 additions and 7 multiplications. The code
optimization package Haggies~\cite{reiter} implements this method of
Horner schemes followed by CSE.

Finding the optimal order of variables for the Horner scheme is still
an open problem for all but the smallest polynomials, which are
studied in Ref.~\cite{kojima}. Different orders may impact the cost of
the resulting code, although no thorough study of this has been made
to the authors' knowledge. Simple algorithms have been proposed in the
literature, such as most-occurring variable first, which results in
the highest decrease of the cost at that particular step. This is also
the order that is used in Haggies.

We studied the results of choosing different orders of variables for
the Horner scheme and discovered that this order greatly affects the
results, sometimes improving the cost by factors up to
two. Unfortunately, most often it is impossible to perform an
exhaustive search through all Horner schemes, since their number
increases as the factorial of the number of variables. Therefore we
have devised a method to find efficient orders by using Monte Carlo
tree search (MCTS)~\cite{chaslot,browne}, a recently proposed search method
from the field of artificial intelligence.

\section{Monte Carlo tree search}
MCTS is a best-first search method that uses random sampling to guide
the traversal of the search tree. It has recently drawn much attention
due to its application in the field of computer Go~\cite{chaslot}, a
classic board game in which computers have traditionally played
weakly. In the past decade the application of MCTS has improved the
playing strength of computers from the level of advanced beginners to
the level of strong amateur players. MCTS has also been applied
successfully in numerous other games and optimization
problems~\cite{browne}.

\begin{figure}[t]
\includegraphics[width=\columnwidth, height=4cm]{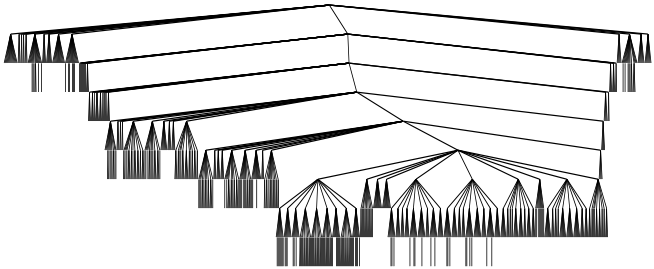}
\caption{The asymmetric search through an MCTS search tree during the
  search for an efficient Horner scheme.}
\label{fig::MCTStree}
\end{figure}

\begin{figure}[t]
\includegraphics[width=\columnwidth]{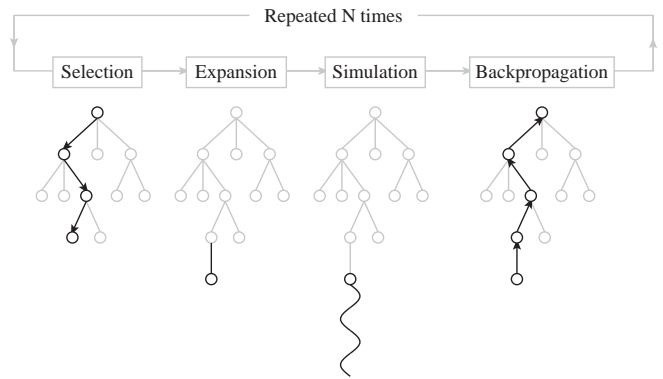}
\caption{The Monte Carlo tree search cycle illustrated. (Figure from
  Ref.~\cite{chaslot})}
\label{fig::MCTSalgo}
\end{figure}

In MCTS the search tree is built in an incremental and asymmetric way,
see Fig.~\ref{fig::MCTStree}. During the search the traversed part of
the search tree is completely in memory. For each node MCTS keeps
track of the number of times it has been visited and the estimated
result of that node. At each step one node is added to the search tree
according to a criterion that tells where most likely better results
can be found. From that node an outcome is sampled and the results of
the node and its parents are updated. This process is illustrated in
Fig.~\ref{fig::MCTSalgo}. In more detail the four steps of the MCTS
cycle are the following.

\textbf{Selection} During the selection step the node which most
urgently needs expansion is selected. Several criteria are proposed,
but the easiest and most-used is the UCT (upper confidence level for
trees) criterion~\cite{kocsis}:
\begin{equation}
UCT_i = \left<x_i\right> + 2 C_p \sqrt{\frac{2\log{n}}{n_i}}.
\label{eqn::UCT}
\end{equation}
Here $\left<x_i\right>$ is the average score of child $i$, $n_i$ is
the number of times child $i$ has been visited and $n$ is the number
of times the node itself has been visited. $C_p$ is a
problem-dependent constant that should be determined
empirically. Starting at the root of the search tree, the
most-promising child according to this criterion is selected and this
selection process is repeated recursively until a node is reached with
unvisited children. The first term of Eqn.~(\ref{eqn::UCT}) biases in
favor of nodes with previous high rewards (exploitation), while the
second term selects nodes that have not been visited much
(exploration). Balancing exploitation versus exploration is essential
for the good performance of MCTS.

\textbf{Expansion} The selection step finishes with a node with
unvisited children. In the expansion step one of these children is
added to the tree.

\textbf{Simulation} In the simulation step a single possible outcome
is simulated starting from the node that has just been added to the
tree. This simulation can consist of generating a complete random
outcome starting from this node or can be some known heuristic for the
search problem. The latter typically works better if specific
knowledge of the problem is available.

\textbf{Backpropagation} In the backpropagation step the results of
the simulation are added to the tree, specifically to the path of
nodes from the newly-added node to the root. Their average results and
visit count are updated.

This MCTS cycle is repeated a fixed number of times or until the
computational resources are exhausted. After that the best found
result is returned.

\begin{algorithm}[b]
\begin{algorithmic}[1]
  \Statex
  \Function{MCTSHorner}{}
    \State{r $\gets$ new node with empty variable order}
    \For{$i \gets 1\dots$NumberOfTreeExpansions}
      \State{$s \gets$ select($r$)}
      \State{$s \gets$ expand($s$)}
      \State{$x \gets$ simulate($s$)}
      \State{backpropagate($s$,$x$)}
    \EndFor
    \State \Return{best optimized expression found}
  \EndFunction
 
  \State{}

  \Function{select}{$s$}
    \While{$s$ is fully expanded}
      \State{$s \gets \argmax\limits_{\text{children }c\text{ of }s} 
        \frac{x(c)}{n(c)} + 2 C_p \sqrt{\frac{2\log(n(s))}{n(c)}}$}
    \EndWhile
    \State \Return{$s$}
  \EndFunction

  \State{}

  \Function{expand}{$s$}
    \State{$o\gets$ variable order of $s$}
    \State{$x\gets$ random variable not in $o$}
    \State{$o\gets$ append($o$,$x$)}
    \State{$c\gets$ new node with variable order $o$}
    \State{add $c$ to children of $s$}  
    \State \Return{c}
  \EndFunction   

  \State{}

  \Function{simulate}{$s$}
    \State{$o\gets$ variable order of $s$}
    \While{$o$ doesn't contain all variables}
      \State{$x\gets$ random variable not in $o$}
      \State{$o\gets$ append($o$,$x$)}
    \EndWhile
    \State{$e\gets$ HornerScheme(expression, $o$)}
    \State{$e\gets$ CommonSubexpressionElimination($e$)}
    \State \Return{$\frac{\text{NumberOfOperations(expression)}}{\text{NumberOfOperations}(e)}$}  
  \EndFunction

  \State{}

  \Function{backpropagate}{$s$,$\delta x$}
    \While{$s\neq$ null}
      \State{$x(s) \gets x(s) + \delta x$}
      \State{$n(s) \gets n(s) + 1$}
      \State{$s\gets$ parent of $s$}
    \EndWhile
  \EndFunction
\end{algorithmic}

\caption{Pseudocode of MCTS Horner}
\label{alg::MCTSHorner}
\end{algorithm}

\section{Efficient Horner schemes}
In the existing code optimization packages that use Horner schemes
combined with CSE, a simple algorithm for the order of the variables
is chosen. Widely used is the \emph{occurrence order}, where the
variables are sorted with respect to the number of occurrences in the
polynomial~\cite{ceberio}. The variable that has the largest number of
occurrences comes first in the order and is the first one used in
Horner's method.

To test whether this algorithm gives efficient Horner schemes we took
a large polynomial with 15 variables, a result from HEP calculations,
and generated a million random orders which were used for Horner's
method followed by CSE. The occurrence order performed quite well,
about a standard deviation above average, but far better orders were
also found. An interesting feature of the orders that led to efficient
schemes also showed up: these orders all shared the same variables in
the trailing part of the order. These are the variables that
eventually show up most often in the common subexpressions. These
common subexpressions abound in the HEP polynomials due to much
structure, such as combinations of coupling constants or dot products
and masses.

Motivated by this observation we use MCTS to determine an order of the
variables that gives efficient Horner schemes. The root of the search
tree represents that no variables are chosen yet. This root node has
$n$ children, with $n$ the number of variables. The other nodes
represent choices for a number of variables in the trailing part of
the order. This number equals the depth of the node in the search
tree. A node at depth $d$ has $n-d$ children: the remaining unchosen
variables.

In the simulation step the incomplete order is completed with the
remaining variables added randomly. This complete order is then used
for Horner's method followed by CSE. The number of operations in this
optimized expression is counted. The selection step uses the UCT
criterion with as score the number of operations in the original
expression divided by the number of operations in the optimized
one. This number increases with better orders and is typically of
${\cal O}(1)$. The constant $C_p$ in Eqn.~(\ref{eqn::UCT}) must
therefore be chosen of that size as well. Pseudocode of MCTS generated
Horner schemes can be found in algorithm~\ref{alg::MCTSHorner}.

\section{Results}

To test the performance of this method an implementation is added to
the computer algebra package {\sc Form}~\cite{form}, which is widely
used for HEP calculations. The results of this method are compared to
a few existing algorithms. For comparison we added to {\sc Form}
optimization routines that use occurrence order Horner schemes
followed by CSE. Furthermore, we compare to the open-source code
optimization package Haggies~\cite{reiter} and the results from the
paper on the hypergraph method based on partial syntactic
factorization~\cite{leiserson}. We also tried the code optimization
routines of Mathematica and Maple, but their results were not of
particular interest. Finally, we present the results of the new code
optimization routines of {\sc Form}, which consist of MCTS generated
Horner schemes followed by greedy optimizations. A detailed description
of this algorithm will be presented in Ref.~\cite{optimize}. Since
this algorithm is an extension of MCTS Horner with CSE, it should
perform better.

\begin{table*}[t]
\centering
\footnotesize

\begin{tabular}{|l|c|c|c|c|c|c|}
\cline{2-7}
\multicolumn{1}{l|}{}& res(7,4) & res(7,5) & res(7,6) & HEP($\sigma$) & HEP($F_{13}$) & HEP($F_{24}$) \\
\hline
%Num. variables      &       13 &       14 &       15 &            15 &            29 &          36  \\
No optimizations    &  29\,163 & 142\,711 & 587\,880 &       47\,424 &   1\,068\,153 & 7\,722\,027  \\
\hline
Occ. Horner + CSE   &   4\,968 &  20\,210 &  71\,262 &        6\,744 &       92\,617 &    401\,530  \\
Haggies             &   7\,540 &  29\,125 & 101\,821 &       13\,214 &      238\,093 & \emph{crash} \\
Hypergraph + CSE    &   4\,905 &  19\,148 &  65\,770 &         ---   &         ---   &       ---    \\
\hline
MCTS + CSE          & $(3.9\pm0.1)\cdot10^3$     & $(1.5\pm0.2)\cdot10^4$   &  $(5.0\pm0.6)\cdot10^4$ 
                    & $(4.3\pm0.3)\cdot10^3$     & $(6.9\pm0.4)\cdot10^4$   &  $(4.4\pm0.2)\cdot10^5$ \\
     $[N=300]$      & [$C_p = 0.03$]             & [$C_p = 0.03$]           &  [$C_p = 0.01$]            
                    & [$C_p = 0.35$]             & [$C_p = 0.03$]           &  [$C_p = 0.01$]         \\
\hline
MCTS + CSE          & $(3.86\pm0.03)\cdot10^3$  & $(1.39\pm0.01)\cdot10^4$  &  $(4.58\pm0.05)\cdot10^4$ 
                    & $4\,114 \pm 14 $          & $(6.6\pm0.2)\cdot10^4$    &  $(3.80\pm0.06)\cdot10^5$ \\
     $[N=1\,000]$   & [$C_p = 0.1$]             & [$C_p = 0.07$]            &  [$C_p = 0.05$]            
                    & [$C_p = 0.75$]            & [$C_p = 0.2$]             &  [$C_p = 0.015$]         \\
\hline
MCTS + CSE          & $(3.84\pm0.01)\cdot10^3$  & $13\,786 \pm 28$          &  $(4.54\pm0.01)\cdot10^4$ 
                    & $ 4\,087 \pm 5 $          & $(6.47\pm0.08)\cdot10^4$  &  $(3.19\pm0.04)\cdot10^5$  \\
     $[N=10\,000]$  & [$C_p = 0.2$]             & [$C_p = 0.2$]             &  [$C_p = 0.15$]            
                    & [$C_p = 1.5$]             & [$C_p = 0.3$]             &  [$C_p = 0.03$]          \\
\hline
MCTS + greedy      & $(3.03\pm0.03)\cdot10^3$   & $(1.09\pm0.01)\cdot10^4$  &  $(3.57\pm0.01)\cdot10^4$ 
                   & $3\,401 \pm 31$            & $(4.63\pm0.09)\cdot10^4$  &  $(1.84\pm0.04)\cdot10^5$    \\
     $[N=10\,000]$ & [$C_p = 0.2$]              & [$C_p = 0.2$]             &  [$C_p = 0.15$]            
                   & [$C_p = 1.5$]              & [$C_p = 0.3$]             &  [$C_p = 0.03$]          \\
\hline
\end{tabular}

\caption{The number of operations in the final expression after
  applying the optimization methods on the polynomials. The results
  for the hypergraph method are from Ref.~\cite{leiserson}; we did not
  have the opportunity to test this method on the HEP
  polynomials. Different MCTS experiments are performed with different
  values for $C_P$, which are determined by trial and error. These are
  present in the table underneath their corresponding results. The
  MCTS results are statistical averages of different samples.}
\label{tbl::results}
\end{table*}

The polynomials to be optimized consist of two sets. The first set of
polynomials (taken from Ref.~\cite{leiserson}) are the resultants of
two polynomials, $\text{res}_x(a(x),b(x))$, with
$a(x)=\sum_{i=0}^ma_ix^i$ and $b(x)=\sum_{i=0}^nb_ix^i$, which is
viewed as a polynomial in the $(m+n+2)$ variables $a_i$ and $b_i$. The
second set consists of a number of large multivariate polynomials
resulting from HEP calculations~\cite{grace}. For the set of
resultants we observed that the variables that are factored out first
in the Horner scheme are critical for the performance of MCTS Horner,
as opposed to the last variables which are important for the HEP
polynomials. This is probably due to many common subexpressions
appearing in the HEP polynomials if the right variables are chosen
last. The resultants don't have that property and are probably more
sensitive to a large decrease in the number of operations due to the
Horner scheme itself. This conjecture must be investigated more
rigorously. In the experiments for the resultants MCTS will search
therefore for the best leading part of the variable order, while for
the HEP polynomials it will search for the best trailing part.

\begin{figure}[b]
\includegraphics[width=\columnwidth]{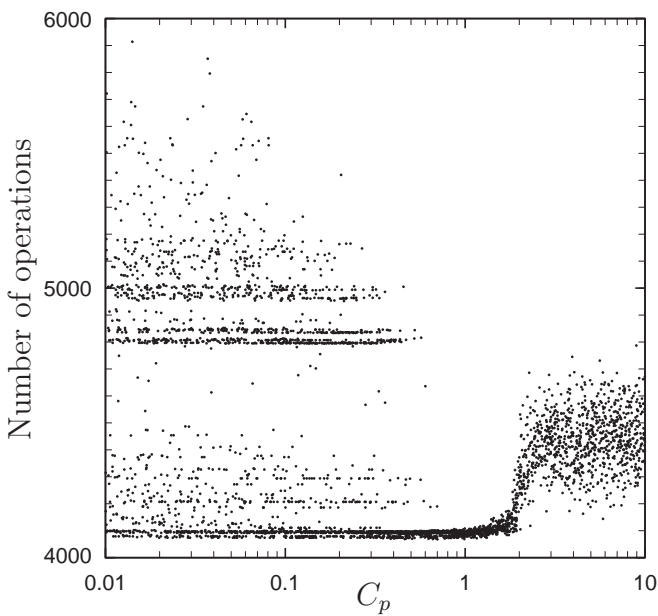}
\caption{A scatter plot of the results of MCTS Horner with CSE with
  $3\,000$ tree expansions for the polynomial HEP($\sigma$) as a
  function of the constant $C_P$ that determines the balance between
  exploitation (small values) and exploration (large values). Shown
  are $4\,000$ randomly chosen values for $C_p$ with their
  corresponding results.}
\label{fig::scatter}
\end{figure}

The results of the optimizations are expressed in the number of
operations in the final expressions and are in
Tab.~\ref{tbl::results}. It is clear that MCTS Horner with CSE beats
the existing algorithms, if the parameters (the
exploitation/exploration constant $C_p$ from Eqn.~(\ref{eqn::UCT}) and
the number of tree expansions $N$) are chosen properly.

\begin{figure}[b]
\includegraphics[width=\columnwidth]{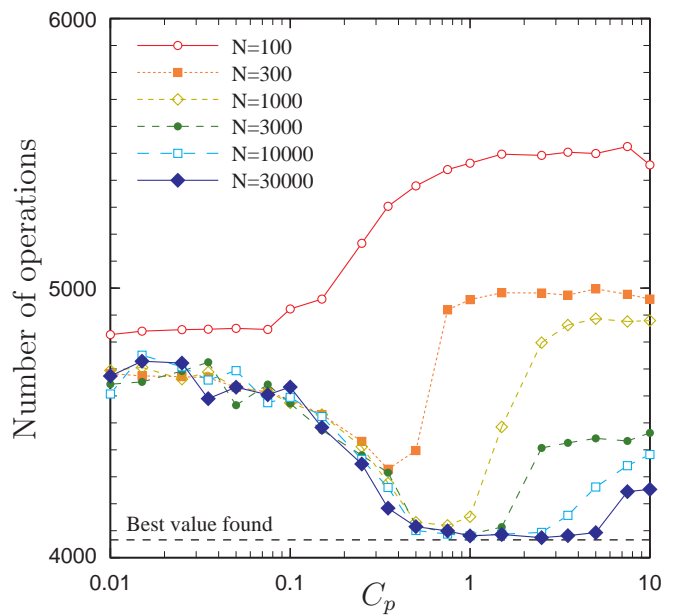}
\caption{The results of MCTS Horner with CSE for the polynomial
  HEP($\sigma$) as function of the exploitation/exploration constant
  $C_p$ and the number of tree expansions $N$. All data points are
  statistical averages over at least 100 samples. The fourth line
  from above (green) consists of the average values per $C_P$ of the
  scatter plot of Fig.~\ref{fig::scatter}. }
\label{fig::constant}
\end{figure}

The effectiveness of MCTS depends heavily on the choice for these
parameters. The results of MCTS with $3\,000$ tree expansions, followed
by CSE, as a function of $C_p$ are in Fig.~\ref{fig::scatter} for a
large polynomial from HEP. For equal values of $C_p$ different
results are produced because of different seeds of the random number
generator. For small values of $C_p$, such that MCTS behaves
exploitively, the method gets trapped in one of the local minima as
can be seen from the different \emph{lines} in the left-hand side of
the figure. For large values of $C_p$, such that MCTS behaves
exploratively, lots of options are considered and no real minimum is
found as can be seen from the \emph{cloud} of points on the right-hand
side. For intermediate values of $C_p\approx1$ MCTS balances well
between exploitation and exploration and finds almost always a Horner
scheme that is very close to the best one known to us.

The results of improving this polynomial for different numbers of tree
expansions are shown in Fig.~\ref{fig::constant}.  For small numbers
of tree expansions it turns out to be good to choose a low value for
the constant $C_p$ (smaller than $0.5$). The search is then mainly
driven by exploitation. MCTS quickly searches deep in the tree, most
likely around a local minimum. This local minimum is explored quite
well, but the global minimum is likely to be missed. With higher
numbers of tree expansions a value for $C_p$ in the range $[0.5;2]$
seems suitable. This gives a good balance between exploring the whole
search tree and exploiting the promising nodes. Really high values of
$C_p$ seem a bad choice in general, since promising nodes are not
exploited anymore. Note that these values hold for this particular
polynomial, and that different polynomials give different optimal
values for $C_p$ and $N$. The different values for $C_p$ in
Tab.~\ref{tbl::results} are determined by trial and error and give
decent results. With some more tuning even better results can probably
be achieved. Automatic tuning of this parameter would be very
convenient and is part of ongoing research.

\begin{table}[t]
\centering
\footnotesize
\begin{tabular}{|l|c|c|}
\hline
Algorithm            & Num. operations & Run time    \\
\hline
No optimizations     &   142\,711   &    ---         \\
\hline
Occurrence Horner + CSE & 20\,210   &       0.29 sec \\ 
Haggies              &    29\,125   &      11.3 sec  \\
Hypergraph + CSE     &    19\,148   &      22.1 sec  \\
\hline

MCTS+CSE $(N\!=\!300, C_P\!=\!0.03) $   & $(1.5\pm0.2)\cdot10^4$   & $47.4$ sec \\ 
MCTS+CSE $(N\!=\!10^3, C_P\!=\!0.07) $  & $(1.39\pm0.01)\cdot10^4$ & $157$ sec \\ 
MCTS+CSE $(N\!=\!10^4, C_p\!=\!0.2) $   & $13\,786 \pm 28$         & $1.4\cdot10^3$ sec \\ 
\hline
MCTS+greedy $(N\!=\!10^4, C_P\!=\!0.2)$ & $(1.09\pm0.01)\cdot10^4$ & $1.4\cdot10^3$ sec\\
\hline
\end{tabular}
\caption{The run times of the various optimization algorithms on the
  polynomial $\text{res}(7,5)$. All tests are performed on 2.4GHz
  Pentium or Xeon CPUs.}
\label{tbl::runtimes}
\end{table}

The different algorithms vary a lot regarding the consumed
computational resources, see Tab.~\ref{tbl::runtimes}. MCTS Horner
with CSE needs considerably more run time than a greedy Horner scheme
with CSE. This makes sense, because MCTS basically does such an
operation per tree expansion. With only few computational resources
available it is better to use a greedy Horner scheme than to use
MCTS. If used it searches through the tree for good schemes for too
short a time and does not find one, therefore resulting in a bad
scheme. Compared to Haggies and the hypergraph method, MCTS with 300
expansions gives slightly longer run times for slightly better
results. When the quality of the polynomial evaluation scheme is of
great importance, it makes sense to spend more time to find a better
evaluation scheme. With more time available, so that large parts of
the search tree can be traversed, MCTS Horner with CSE or greedy
optimizations outperforms all other methods considerably.

\section{Discussion}

Polynomials are fundamental mathematical objects, naturally occurring
at many places in mathematics and science. Efficient evaluation of
polynomials is of great importance to many application areas. Horner's
method is a simple approach straight out of undergraduate algorithms
textbooks. For something so basic, it is remarkable that over the
years so little improvement has been made in finding more efficient
evaluation schemes.

We improve on the traditional multivariate Horner schemes, where the
variable order is fixed by a simple procedure, by employing MCTS. By
statistically sampling the different variable orders it is designed to
balance exploitation of known good schemes while not forgetting to
explore unknown schemes as well.

The basic multivariate Horner schemes generated with most-occurring
variables ordered first will quickly yield schemes that may be
efficient enough for many applications. More demanding domains, where
the expressions are large and/or evaluated many times, need better
evaluation schemes. For these applications it pays to invest the time
to generate them. MCTS is suitable for these types of
applications. Evaluating large expressions for Feynman diagrams in HEP
is one such domain, where large expressions have to be evaluated many
times doing Monte Carlo integration~\cite{formcalc,grace,madgraph,gosam}.

For all examined polynomials MCTS Horner, followed by CSE, generated
evaluation schemes that were better than any other algorithm that we
tried. A further analysis of the performance is ongoing research. This
includes sensitivity analysis to different parameters
(exploitation/exploration constant, number of tree expansions),
dependence on the polynomial (size, number of variables, use of
heading or trailing part of the variable order), automatic tuning of
the parameters to the polynomial, better criteria for the selection
step, possible heuristics for the simulation step (such as using the
occurrence order instead of random completion), and the
parallelization of the algorithm.

\end{document}